\documentclass[12pt]{article}
\usepackage{amsmath}
\usepackage{graphicx}
\usepackage{amssymb}

\newcommand{\be}{\begin{equation}}
\newcommand{\ee}{\end{equation}}
\newcommand{\bea}{\begin{eqnarray}}
\newcommand{\eea}{\end{eqnarray}}
\newcommand{\bs}{\boldsymbol}

\begin{document}

\bigskip
\begin{titlepage}

\begin{flushright}
NORDITA-2007-11\\
hep-th/yymmnnn
\end{flushright}

\vspace{1cm}

\begin{center}
{\LARGE Inflation Assisted by Heterotic Axions}

\end{center}
\vspace{3mm}

\begin{center}

{\bf{Martin E. Olsson} \\
\vspace{5mm}
}

\it{NORDITA, Roslagstullsbacken 23\\ 
SE - 106 91 Stockholm, Sweden}

\vspace{5mm}

{\tt
\bf{molsson@nordita.dk}
}

\end{center}
\vspace{5mm}

\begin{center}
{\large \bf Abstract}
\end{center}
\noindent
We explore the possibility of obtaining inflation in weakly coupled heterotic string theory, where the model dependent axions are responsible for driving inflation. This model can be considered as a certain extrapolation of $m^{2}\phi^{2}$-inflation, and is an attempt to explicitly realize the so called N-flation proposal in string theory. The instanton generated potential for the axions essentially has two parameters; a natural mass scale $M$ and the string coupling $g_{s}$. For isotropic compactifications leading to of order $\mathcal{O} (10^4)$ axions in the four dimensional spectrum we find that with $(M,\mbox{ } g_{s})\simeq(M_{GUT},\mbox{ } 0.5)$ the observed temperature fluctuations in the CMB are correctly reproduced. We assume an initially random distribution for the vevs of the axions. The spectral index, $n_{s}$, is generically more red than for $m^{2}\phi^{2}$-inflation. The greater the vevs, the more red the spectral index becomes. Allowing for a wide range of vevs $55$ $e$-foldings from the end of inflation, we find $0.946\lesssim n_{s} \lesssim 0.962$. The tensor-to-scalar ratio, $r$, is more sensitive to the vevs, but typically smaller than in $m^{2}\phi^{2}$-inflation. Furthermore, in the regime where the leading order theory is valid, $r$ is bounded by $r < 0.10$. The spectral index and the tensor-to-scalar ratio are correlated. For example, $n_{s}\simeq 0.951$ corresponds to $r\simeq 0.036$.

\vfill
\begin{flushleft}
\bigskip
February 2007
\end{flushleft}
\end{titlepage}
\newpage

\section{Introduction and summary}
\bigskip

With the growing evidence for inflation in the early universe it is important to investigate if, and under what conditions, inflation can be realized in string theory. Ideally, one would like to obtain inflation in string theory without fine-tuning and in a setting where higher order corrections are under controll.

Typical models of inflation are large field models, where the inflaton $\phi$ must roll over super-Planckian distances in order to generate an appropriate number of $e$-foldings. From an UV-perspective this would typically generate large corrections due to terms such as $\phi^2/M_{Pl}^2$, rendering the first order approximation invalid. Furthermore, for single fields, the potential must be flat enough in order to give substantial slow-roll, suggesting that some amount of tuning is required.

One effective model which has these features, but is nevertheless a good fit to recent CMB data \cite{Spergel:2006hy}, is the single field $m^2 \phi^2$-model of chaotic inflation. While it appears unlikely that this model, as it stands, can be embedded into string theory in a controlled way, it may be a good approximation to a more general multi-field potential with a quadratic minima. (Indeed, almost any potential with a nontrivial minima is quadratic close to that point.) The feature of using many fields is that while a single field only rolls over sub-Planckian distances, the collective field may, if the number of fields is sufficiently large, roll over super-Planckian distances.

Furthermore, even though the potential for a single field is too steep in order to inflate, the presence of many fields combine to generate enough friction for the individual fields to roll sufficiently slow. In this way, fine tuning of the potential is replaced by the requirement of a large number of inflaton fields.

Finally, the presence of a large number of scalar fields is quite generic in string theory compactifications. It would be nice to use them for something.

An explicit example of multiple inflaton fields, inspired by string theory, is inflation driven by many compact scalars, so called axions. This is the N-flation model of \cite{Dimopoulos:2005ac}, and is an example of \emph{assisted} inflation \cite{Liddle:1998jc}. Aspects of N-flation were further examined in \cite{Easther:2005zr,Kim:2006ys,Piao:2006nm,Singh:2006yy,Kim:2006te,Battefeld:2006sz,Seery:2006js,Gong:2006zp}. Assisted inflation has been investigated in many papers \cite{Kanti:1999vt,Malik:1998gy,Copeland:1999cs,Kanti:1999ie,Kaloper:1999gm,Jokinen:2004bp,Brandenberger:2003zk,Piao:2002vf}. An M-theory model of assisted inflation is \cite{Becker:2005sg}, which was further analyzed in \cite{Ashoorioon:2006wc}.

The purpose of this note is to examine whether N-flation can be concretely realized in a specific string theory setting. It is found that one promising candidate is weakly coupled heterotic string theory, with inflation driven by the model dependent axions.

The reason this setting is particularly appropriate for this purpose is that the instanton action (which exponentially suppresses the axion potential) corresponding to these axions, is small enough to generate a sufficiently large potential. The instanton in question is a world-sheet instanton, which at weak coupling results in an action which is at least around an order of magnitude smaller than instanton actions due to D-brane instantons, appearing for instance in type II string theory. The difference in magnitude can be understood from the way the effects depend on the string coupling constant \cite{Svrcek:2006yi,Svrcek:2006hf}.

For degenerate axion potentials and small field values, this model mimics the $m^2\phi^2$-model of chaotic inflation \cite{Dimopoulos:2005ac}\footnote{The field $\phi$ here represent either the single inflaton or, in the multi-field case, the collective field in chaotic inflation.}. The reason is that the axion potentials are quadratic close to the origin. With the more plausible assumption of initially \emph{random} axion vacuum expectation values (vevs) \cite{Kim:2006ys}, the model can be considered as a certain extrapolation of $m^{2}\phi^{2}$-inflation, that naturally can be realized in string theory. With this assumption for the initial conditions, the outcome for the observables computed will be modified as compared to chaotic inflation.

Because of the embedding into string theory the parameter space has significantly shrunk compared to effective modeling. We show that withing this window it is possible to reproduce the observed temperature fluctuations in the CMB. Specifically, assuming degenerate axion potentials, the complete potential depends essentially only on two parameters; a natural mass scale $M$ and the string coupling $g_{s}$. It also depends on the number of axions $N$, but this can be swapped for the number of $e$-foldings $N_{e}$. In order to get a sufficient amount of inflation, we need of order $\mathcal{O}(10^4)$ axions in the four-dimensional spectrum. With $(M, g_{s})\simeq (M_{GUT},\mbox{ } 0.5)$ and initially random axion vevs, the observed CMB fluctuations are correctly reproduced.

There are a number of possible higher order corrections which may invalidate the leading order result. These are string loop, world-sheet and multi-instanton corrections. Keeping these corrections under controll puts a further bound on the parameter space, in particular on the number of axions allowed. Values of parameters used in the leading order theory, such as the string coupling and the number of axions, are well within the allowed range as long as the axion vevs did not take too small values when scales observed today were produced.

Observables such as the scalar spectral index $n_{s}$ and the tensor-to-scalar ratio $r$ only depend on the initial conditions of the axion vevs. This is true for isotropic compactifications, which here implies degenerate axion potentials. The dynamics of N-flation with non-degenerate potentials and small initial axion vevs was studied in \cite{Easther:2005zr}.

The spectral index, which is not very sensitive to the initial conditions, is well within current observational bounds for a wide range of vevs at about 55 $e$-foldings from the end of inflation, and is typically more red than for the $m^2\phi^2$-model. In the regime where the leading order theory is valid, the spectral index takes values in the interval $0.946 \lesssim n_{s} \lesssim 0.962$. The precise value is uniquely given by the distribution of the axion vevs. The running of the spectral index is completely negligible.

The tensor-to-scalar ratio is more sensitive to the initial vevs. It is typically smaller than the corresponding result for chaotic inflation. Interestingly, requiring validity of the leading order theory puts the sharp upper bound $r<0.10$ on the tensor-to-scalar ratio. However, $r$ is still large enough to be detected by upcoming observations. For example, a detection of $n_{s}\simeq 0.951$ corresponds here to $r\simeq 0.036$.

The organization of the paper is as follows: In section \ref{sec:axioninfl} we begin by discussing the axion and its potential in some detail, with particular emphasis on the model dependent heterotic axions. Then we work out the resulting inflationary dynamics and compare to CMB observations, showing that the weakly coupled heterotic string is a promising setting for N-flation. We give conditions for keeping higher order corrections as well as other effects under control. In section \ref{sec:cosmobs} we extract various cosmological observables from this model. In the concluding section \ref{sec:conclusions}, we, amongst other things, discuss effects that may modify the results obtained in the simple treatment presented here.

\bigskip
\section{Inflation from string theory axions} \label{sec:axioninfl}
\bigskip

Axions are scalars with a perturbative global symmetry $a \rightarrow a+c$, where $a$ is the axion and $c$ is an arbitrary constant. Nonperturbatively, this symmetry is broken by instanton effects down to a discrete symmetry, generating a mass and a potential for the axion. In this section we show that in weakly coupled heterotic string theory, the instanton generated potential can inflate and produce CMB fluctuations consistent with observations.

\subsection{The axion potential} \label{sec:axionpot}

In string theory, effects that break the shift symmetry for the axions arise at nonperturbative level. The symmetry breaking is due to instantons, that have a specific origin in a given string theory model. The resulting potential is exponentially suppressed by the instanton action and obtains the form
\be
V(a)=M^4e^{-S}\left[1-\cos\left(\frac{a}{F_{a}}\right)\right],
\ee
where $M$ is some natural mass scale, $S$ the instanton action and $F_{a}$ the axion decay constant. The trigonometric dependence reflects the residual discrete shift symmetry $a \rightarrow a + 2\pi n F_{a}$. For brevity, we will from now on, unless stated otherwise, use the rescaled notation $a \rightarrow aF_{a}$. In this notation, the axion field $a$ is dimensionless and takes on values $-\pi \leq a \leq \pi$.

The total potential with $N$ axions is simply given by
\be
V=\sum_{i=1}^{N}{V(a_{i})},
\ee
if we assume no cross-couplings among the axions\footnote{This is so far an unjustified assumption which we will come back to later.}. Furthermore, when the different instanton actions corresponding to the individual axions are of roughly the same size, we can simply write down
\be \label{eq:inflpot}
V=\sum_{i=1}^{N}{V(a_{i})}\simeq NM^4e^{-S}\frac{1}{N}\sum_{i=1}^{N}{\left( 1-\cos a_{i}\right)}.
\ee
As we will see later, the assumption of similar instanton actions translates into requiring a roughly isotropic compactification.

Let us now make some remarks on the parameters entering the axion potential (\ref{eq:inflpot}), i.e. the number of axions $N$, the natural mass scale $M$, the instanton action $S$ and, implicitly, the axion decay constant $F_{a}$. The number of axions $N$ controls the number of $e$-foldings, and we will take it to be as large as is required in order to get a sufficient number of $e$-foldings. More detailed studies will decide if the required number is attainable in given compactifications. Here we will only verify that it is below the regime where higher order correction terms become important. 

For the mass scale $M$ it is, a priori, natural to assume that it is no larger than the reduced Planck mass and no smaller than the string or GUT scale. The instanton action $S$ and the axion decay constant $F_{a}$ are given by a direct calculation in a given string theory. This has been done in various string theory settings and most recently in \cite{Svrcek:2006yi}. We will briefly review these calculations in the following section, for the specific case of heterotic string theory. This will also serve to explain why the weakly coupled heterotic string is a particularly promising setting for realizing N-flation in string theory.

\subsubsection*{\it{Table of sums}}

In the following sections we will encounter many different sums, originating from (\ref{eq:inflpot}), corresponding to various observables. For notational purpose, we collect these sums here for reference:
\begin{align} 
\bs{s_{1}}&=\frac{1}{N}\sum_{i=1}^{N}{\left( 1-\cos a_{i}\right)} \label{eq:sum1} \\
\bs{s_{2}}&=\frac{1}{N}\sum_{i=1}^{N}{\left[\ln 2+\ln \left(\frac{1-\cos a_{i}}{\sin^{2} a_{i}}\right)\right]} \label{eq:sum2} \\
\bs{s_{3}}&=\frac{1}{N}\sum_{i=1}^{N}{\sin^{2}a_{i}} \label{eq:sum3} \\
\bs{s_{4}}&=\frac{1}{N}\sum_{i=1}^{N}{\left(\frac{1-\cos a_{i}}{\sin a_{i}}\right)^{2}} \label{eq:sum4} \\
\bs{s_{5}}&=\frac{1}{N}\sum_{i=1}^{N}{\frac{\cos a_{i}(1-\cos a_{i})^{2}}{\sin^{2}a_{i}}} \label{eq:sum5},
\end{align}

Later, when we will calculate observables, we need to evaluate these sums. Note that in doing this, the actual value of $N$ is not important (as long as $N$ is large enough) since we sum over $N$ terms and divide by $N$ in the end. What matters is how the axions are distributed in their respective potentials.

\subsection{Heterotic axions}
We first compactify the theory on a six-manifold $Z$ with volume $V_{Z}$, so that we end up with the product space $\mathbb{R}^{1,3}\times Z$. The axions in weakly coupled heterotic string theory come from the zero modes of the NS-NS $B$-field. The components $B_{\mu \nu}$ with $\mu$ and $\nu$ tangent to the noncompact 4$d$ space can be dualized to make the so called model independent axion. Components tangent to the compact space $Z$ instead give rise to the model dependent axions. Instantons that break the shift-symmetry of these latter axions are world-sheet instantons wrapping two-cycles within $Z$. With the fundamental string tension $\tau_{f}=1/(2\pi \alpha')$ and $R^2$ the size of a given two-cycle, the instanton action is given by
\be \label{eq:instact1}
S=\tau_{f} R^2=\frac{R^2}{2\pi \alpha '}=\frac{2\pi R^2}{l_{s}^2},
\ee 
where we used the convention $l_{s}=2\pi \sqrt{\alpha '}$. $R^2$ is bounded by the size of the compactification volume $V_{Z}$, and, for a reasonably isotropic $Z$, is roughly given by $R^2 \simeq V_{Z}^{1/3}$. In order to find an expression for the compactification volume, we dimensionally reduce the ten-dimensional heterotic gauge-gravity action \cite{Polchinski:1998rr}
\be
S_{10}=\frac{2\pi}{g_{s}^{2}l_{s}^{8}}\int{d^{10}x\sqrt{-g}R}-\frac{1}{8\pi g_{s}^{2}l_{s}^{6}}\int{d^{10}x\mbox{ }\text{tr}\mbox{ }F_{MN} F^{MN}}
\ee
down to four dimensions by compactifying on $Z$
\be
S_{4}=\frac{2\pi V_{Z}}{g_{s}^{2}l_{s}^{8}}\int{d^{4}x\sqrt{-g}R}-\frac{V_{Z}}{16\pi g_{s}^{2}l_{s}^{6}}\int{d^{4}x\mbox{ }\text{tr}\mbox{ }F_{\mu \nu} F^{\mu \nu}},
\ee
from which we can read off
\be \label{eq:planckgaugeexp}
M_{Pl}^{2}=\frac{4\pi V_{Z}}{g_{s}^{2}l_{s}^{8}}\mbox{ }, \quad g_{YM}^{2}=\frac{4\pi g_{s}^{2}l_{s}^{6}}{V_{Z}}\mbox{ }.
\ee
In terms of the unified gauge coupling $\alpha_{G}=g_{YM}^{2}/4\pi$ we have
\be \label{eq:alphag}
\alpha_{G}=\frac{g_{s}^{2}l_{s}^{6}}{V_{Z}}.
\ee
This immediately gives the volume of the compact manifold $V_{Z}$ in terms of gauge and string theory parameters
\be \label{eq:hetvol}
V_{Z}=\frac{g_{s}^{2}l_{s}^{6}}{\alpha_{G}}.
\ee
Using this expression for the volume of the compact manifold, the instanton action (\ref{eq:instact1}), with $R^{2}\simeq V_{Z}^{1/3}$, can now be written as
\be
S=\frac{2\pi R^2}{l_{s}^2}\simeq \frac{2\pi V_{Z}^{1/3}}{l_{s}^2}=\frac{2\pi g_{s}^{2/3}}{\alpha_{G}^{1/3}}.
\ee 
Finally, using the ``phenomenological" value of the unified gauge coupling $\alpha_{G}=1/25$, we get
\be \label{eq:instact2}
S\simeq \frac{2\pi g_{s}^{2/3}}{\alpha_{G}^{1/3}} = 18.3 g_{s}^{2/3}. 
\ee

There is an upper and a lower bound on the string coupling $g_{s}$. For the weakly coupled perturbative description to be valid we require $g_{s} \lesssim 1$. We will discuss this condition in more detail later when we consider higher order correction in the presence of a large number of axions. The lower bound can be deduced from the expression of the compactification volume (\ref{eq:hetvol}). $V_{Z}$ cannot be smaller than the string scale, i.e. $V_{Z}\geq l_{s}^{6}$. Otherwise one should make a T-duality transformation and work from there. With $\alpha_{G}=1/25$, the expression for the compactification volume (\ref{eq:hetvol}) gives the lower bound $g_{s} \gtrsim 0.2$. This together with the natural assumption we made for the mass scale $M$ in (\ref{eq:inflpot}) (i.e. that it should be no larger than the reduced Planck scale and no smaller than the GUT scale), provides a rather narrow window for the axion potential.

Note that the volume of the compact manifold (\ref{eq:hetvol}) is close to the string scale. This in turn implies that it cannot be very anisotropic, i.e. the size of the internal cycles cannot be too different. If they were, and say one cycle is large, then, in order to end up with the total volume (\ref{eq:hetvol}), at least one other cycle must be small. But the string scale sets the limit for how small a cycle can be, which therefore, in this case, sets a restrictive bound on how anisotropic the compact manifold can be. As mentioned, we consider only isotropic manifolds, which in view of this remark is a plausible assumption. Since the sizes of the two-cycles determines the instanton actions, the assumption of an isotropic compact manifold allows us to make simplifications such as the one made in (\ref{eq:inflpot}).

We also need an expression for the axion decay constant $F_{a}$, which we only quote here. It can be shown \cite{Svrcek:2006yi} that by dualizing $H$ to $a$ and by dimensionally reducing the kinetic term $-2\pi/g_{s}^{2}l_{s}^{4}\int{H_{MNP}H^{MNP}}$ to four dimensions, one finds
\be
S_{kin}=-\int{d^{4}x \frac{F_{a}^{2}}{2}\partial_{\mu}a\partial^{\mu}a}
\ee
where
\be \label{eq:axiondec}
F_{a}=\frac{xR}{\sqrt{2\pi}g_{s}l_{s}^{2}}
\ee
with again $R^{2}$ the size of a two-cycle in $Z$, and $x$ a constant of order one depending on the details of the compactification manifold $Z$. We assume an isotropic $Z$, so that all axions have the same value of the decay constant, i.e. the same value of the constant $x$ in (\ref{eq:axiondec}) and size $R^{2}$ of the two-cycles. From now on we also take $x=1$. This latter restriction only affects the number of axions we will need, which is a number we do not know how to calculate anyway.

Note that the axion decay constant (\ref{eq:axiondec}) can be written in terms of the instanton action (\ref{eq:instact1}) by using the expression for the Planck mass (\ref{eq:planckgaugeexp})
\be \label{eq:instfrel}
F_{a}=\frac{M_{Pl}}{S}\sqrt{\frac{R^{6}}{2V_{Z}}}\simeq \frac{M_{Pl}}{\sqrt{2}S},
\ee 
where we used $R^{2}\simeq V_{Z}^{1/3}$. The axion decay constant is thus generically smaller than the reduced Planck mass. Since the period of the (dimensionfull) axion $a$ is $2\pi F_{a}$, this in turn implies that a single axion never rolls over super-Planckian distances.

As a last comment here, we note that the instanton action (\ref{eq:instact2}) is quite {\emph{small}} compared to instantons in other string theory settings (see for example \cite{Svrcek:2006yi,Svrcek:2006hf}). There is an intuitive reason for that. The instanton action is given by $S=\tau V_{X}$, where $\tau$ is the tension of the object giving rise to the instanton, and $V_{X}$ is the volume it wraps. In heterotic string theory the object in question is the fundamental string, and the volume a two-cycle within the compactification volume $V_{Z}$. We have seen that this volume is in general small, and decreases with the string coupling. Furthermore, the tension is independent of the string coupling, unlike the tension for D-branes or NS5-branes for example, where the tension increases with decreasing string coupling. The relative smallness of the instanton action is important since the axion potential is exponentially suppressed by this action. Too much suppression gives a potential which is far too small in order to generate inflation. Thus, the fact that the instanton action corresponding to the model dependent axions in heterotic string theory is substantially smaller than other stringy instanton actions \cite{Svrcek:2006yi,Svrcek:2006hf}, is the reason we consider this theory here.

\subsection{Inflationary dynamics and matching to observed CMB fluctuations}

In this section we will discuss the inflationary dynamics of the potential (\ref{eq:inflpot}). With a random distribution of the initial axion vevs, the axion potential (\ref{eq:inflpot}) has essentially two parameters, the mass scale $M$ and, by (\ref{eq:instact2}), the string coupling $g_{s}$. As discussed in the previous section these parameters have a lower and an upper bound. In order for (\ref{eq:inflpot}) to be a phenomenologically acceptable inflationary potential, it must, within this parameter range, reproduce the observed amplitude in CMB density fluctuations. For definiteness, we assume throughout that scales measured today were produced $55$ $e$-foldings from the end of inflation.\footnote{The reason we do not use the more standard number 60 is that the number of $e$-foldings from the end of inflation, when scales observed today were produced, decrease with decreasing reheat temperature \cite{Liddle:2000cg}. Since we are dealing with axions the leading coupling to other fields is suppressed, resulting in a relatively low reheat temperature, at about $10^7$ TeV \cite{Dimopoulos:2005ac}.}

We begin by obtaining an expression for the number of $e$-foldings $N_{e}$ in terms of the number of axions $N$. The equation of motion for a single axion is
\be \label{eq:eqom}
\ddot{a}_{i}+3H\dot{a}_{i}+V_{,i}=0,
\ee
where $H$ is the Hubble parameter, and the derivative of the potential is with respect to the (dimensionfull) axion. The power of assisted inflation is now that while the potential for a single field is too steep in order to inflate, the presence of many fields combine to generate enough friction for the individual fields to roll sufficiently slow. So, when the friction term $3H\dot{a}_{i}$ is large the system is in a slow-roll regime. We can then ignore the acceleration term in (\ref{eq:eqom}) and find
\bea
H^{2}=\frac{V}{3M_{Pl}^{2}}, \label{eq:dynamics1}\\
3H\dot{a}_{i}+V_{,i}=0. \label{eq:dynamics2}
\eea
Inflation lasts for as long as the slow-roll conditions \cite{Lyth:1998xn}
\be
\epsilon=\frac{M_{Pl}^{2}}{2}\left(\frac{V_{,i}}{V}\right)^{2}\ll 1, \qquad \eta=M_{Pl}^{2}\frac{V_{,ii}}{V}\ll 1, 
\ee
are satisfied. It turns out that the second of these two conditions is the most restrictive one in this multi-field case. We use this condition to make a rough estimate of the axion distribution at the end of inflation. With the potential (\ref{eq:inflpot}) we find
\be
\eta=M_{Pl}^{2}\frac{V_{,ii}}{V}=\frac{M_{Pl}^{2}}{F_{a}^{2}}\frac{1}{N\bs{s_{1}}},
\ee
where the sum $\bs{s_{1}}$ was defined in (\ref{eq:sum1}). Using (\ref{eq:instfrel}) we now find the following condition for the axion distribution
\be
\bs{s_{1}}\gtrsim\frac{S^{2}}{N},
\ee
for the potential to still inflate. As we soon will see, the instanton action (\ref{eq:instact2}) is $S\simeq 10$ and we need at least $10^{4}$ axions. We then find that inflation lasts until $\bs{s_{1}}\simeq 1/100$. This means that the average vev at the end of inflation is $\bar{a}\ll 1$. We will use this fact when we now calculate the number of $e$-foldings. 

The number of $e$-foldings is defined by
\be
N_{e}=\int{Hdt},
\ee
which from (\ref{eq:dynamics1}) and (\ref{eq:dynamics2}) becomes
\be
N_{e}=-\frac{1}{M_{Pl}^{2}}\sum_{i}\int{\frac{V_{i}}{V_{,i}}da_{i}}.
\ee
Plugging in the expression for the potential (\ref{eq:inflpot}) we find
\begin{align}
N_{e}&=-\frac{F_{a}^{2}}{M_{Pl}^{2}}\sum_{i}\int_{a_{i}}^{a_{i}^{end}}{\frac{1-\cos a_{i}}{\sin a_{i}}da_{i}} \simeq\frac{F_{a}^{2}}{M_{Pl}^{2}}\sum_{i=1}^{N}{\left[\ln 2+\ln \left(\frac{1-\cos a_{i}}{\sin^{2} a_{i}}\right)\right]} =\frac{F_{a}^{2}N}{M_{Pl}^{2}}\bs{s_{2}}, \label{eq:efolds}
\end{align}
where $\bs{s_{2}}$ was defined in (\ref{eq:sum2}), and we have taken $a_{i}^{end}\ll 1$ for all axions. We immediately find from (\ref{eq:efolds}) that the number of axions required in order to obtain $N_{e}$ $e$-foldings is given by
\be \label{eq:NvsNe}
N=\frac{N_{e}M_{Pl}^{2}}{F_{a}^{2}\bs{s_{2}}}\simeq\frac{2N_{e}S^{2}}{\bs{s_{2}}},
\ee
where we used (\ref{eq:instfrel}) to obtain the last expression. We can now estimate the number of axions required. With $N_{e}=55$, $S\simeq 10$ and a distribution of axions in the interval $0<a_{i}<2-3$, we get $N\simeq (1-3)\times 10^{4}$. This may seem like quite a large number, but there are known string theory compactifications leading to of order $10^{3-4}$ axions in the four dimensional spectrum. Furthermore, it follows from F-theory compactifications that there are string models with up to $10^{5-6}$ axions \cite{Candelas:1997eh}.

Note that the smaller the axion vevs are, the smaller $\bs{s_{2}}$ becomes. Hence more axions are needed in order to get the required number of $e$-foldings. As we will see in the next section, requiring validity of the leading order results imposes an upper bound on the number of axions. This in turn, by (\ref{eq:NvsNe}), sets a limit to how small the axion vevs can be in order to obtain enough inflation. 

We now turn to the scalar power spectrum $P_{\mathcal{R}}$. It is given by \cite{Lyth:1998xn}\footnote{The amplitude of the primordial density perturbations is proportional to $P_{\mathcal{R}}^{1/2}$.}
\be \label{eq:defscalarpower}
P_{\mathcal{R}}=\frac{V}{12\pi^{2}M_{Pl}^{6}}\sum_{i}^{N}\left(\frac{V_{i}}{V_{,i}}\right)^{2},
\ee
which with the potential (\ref{eq:inflpot}) becomes
\be \label{eq:scalarpower}
P_{\mathcal{R}}=\frac{N_{e}^{2}S^2e^{-S}}{6\pi^{2}}\left(\frac{M}{M_{Pl}}\right)^{4}\frac{\bs{s_{4}}\bs{s_{1}}}{\bs{s_{2}}^{2}},
\ee
where we used (\ref{eq:NvsNe}) in order to replace $N$ by $N_{e}$.

Now let us check if (\ref{eq:scalarpower}) can reproduce the observed power spectrum within the parameter range discussed in the previous section. First note that the combination of the sums is of order $\mathcal{O} (1)$. Hence, given the number of $e$-foldings, the value of (\ref{eq:scalarpower}) is essentially given by the instanton action $S$ and the natural mass scale $M$. $S$ in turn is, by (\ref{eq:instact2}), given by the value of the string coupling $g_{s}$. As mentioned at the beginning of this section, we take $N_{e}=55$. For definiteness, we take the axions to be equidistantly distributed in the interval $-3<a_{i}<3$, which gives for the combination of the sums\footnote{At the other end, where the axions are far down the potential, the combination of the sums is about a factor of 3 smaller. Thus, for the purpose of finding if this model can match the observed power spectrum to the right order of magnitude, the actual distribution is not very important.} $\bs{s_{4}}\bs{s_{1}}/\bs{s_{2}}^{2} \simeq 6.5$.

WMAP measures \cite{Spergel:2003cb} $P_{\mathcal{R}}^{obs}=2.3 \times 10^{-9}$. To get the right order of magnitude for the calculated power spectrum, (\ref{eq:scalarpower}), with the mass scale given by the reduced Planck mass $M=M_{Pl}=2.4\times 10^{18}$ GeV, we need a string coupling $g_{s}\simeq 2.4$, which is clearly unacceptable. If instead we take the heterotic string scale\footnote{This can be deduced from Eqs.~(\ref{eq:planckgaugeexp}) and (\ref{eq:alphag}) using $\alpha_{GUT}=1/25$.} $M=M_{s}\equiv l_{s}^{-1}=1.3 \times 10^{17}$ GeV, we need $g_{s}\simeq 1.1$, which is still slightly too large. With the GUT scale $M=M_{GUT}\simeq 2 \times 10^{16}$ GeV, we need $g_{s}\simeq 0.5$ in order to match to the observed power spectrum $P_{\mathcal{R}}^{obs}$, which appears to be acceptable.

With these parameter values, we can now compare to other results in the literature \cite{Dimopoulos:2005ac,Easther:2005zr} by computing the typical axion mass $m_{a}$
\be
m_{a}=V_{,aa}^{1/2}\simeq \frac{M_{GUT}^{2}e^{-S/2}}{F_{a}}\simeq \frac{\sqrt{2}SM_{GUT}^{2}e^{-S/2}}{M_{Pl}}\simeq 10^{10}\mbox{ } \text{TeV},
\ee
in agreement with the results of \cite{Dimopoulos:2005ac,Easther:2005zr}, where instead the axion mass was consider as a free parameter, to be fixed by requiring agreement with the observed power spectrum.

The value $g_{s}\simeq 0.5$ for the string coupling is naively acceptable. However, the presence of many axions might enhance the strength of radiative corrections, so the issue is not clear cut. Whether $g_{s}\simeq 0.5$ is small enough will be explored in the upcoming section, with an affirmative result.

As for the mass scale, rather than the unification scale, it is usually assumed that the relevant scale is the reduced Planck mass or the string scale. On the other hand, the GUT scale is only about a factor $2\pi$ smaller than the string scale. A factor which, for instance, might result from details of dimensional reduction down to the effective $4d$ theory.

Note that with $M=M_{GUT}$ and $g_{s}\simeq 0.5$ the potential (\ref{eq:inflpot}) is of order $V\sim (M_{GUT})^{4}$. It is then tempting to interpret the symmetry breaking scale for the axions (or, put differently, the scale of instanton physics) at $M=M_{GUT}$, as the origin for why inflation appears to be tied to physics at the unification scale. In any case, it would be useful to explicitly obtain this mass scale in heterotic string theory.

In conclusion: Of order $\mathcal{O} (10^{4})$ axions are required in order to generate a sufficient amount of inflation. Furthermore, with $M\simeq M_{GUT}$ and $g_{s}\simeq 0.5$, the temperature fluctuations in the CMB are correctly reproduced.

\subsection{Higher order corrections} \label{sec:corrections}

There are various types of higher order corrections which might invalidate the first order results. These are terms coming from $\alpha'$- as well as $g_{s}$-expansions. Furthermore, we must also justify the assumption of negligible cross-couplings among the axions. We treat these effects in turn, following \cite{Dimopoulos:2005ac}.

\begin{itemize}

\item The first order world-sheet correction to the Planck mass is given by
\be \label{eq:alphacorr}
\frac{\delta_{\alpha'}M_{Pl}^{2}}{M_{Pl}^{2}}\simeq \pm \frac{N M_{UV}^{2}}{16 \pi^{2} M_{Pl}^{2}},
\ee
where $N$ is the number of axions and $M_{UV}$ is a high energy cut-off scale, which we take to be the string scale $l_{s}^{-1}$. Using (\ref{eq:planckgaugeexp}) and (\ref{eq:alphag}) the string scale squared is given by
\be \label{eq:stringscale}
l_{s}^{-2}=\frac{ \alpha_{G}M_{Pl}^{2}}{4\pi}.
\ee
Requiring $\left|\frac{\delta_{\alpha'}M_{Pl}^{2}}{M_{Pl}^{2}}\right| < 1$, then gives the condition
\be
\left|\frac{\delta_{\alpha'}M_{Pl}^{2}}{M_{Pl}^{2}}\right|\simeq  \frac{N l_{s}^{-2}}{16 \pi^{2} M_{Pl}^{2}}=\frac{\alpha_{G}N}{(4\pi)^{3}}<1,
\ee
so, in order to trust the leading order result, the number of axions must satisfy
\be \label{eq:boundaxions}
N<\frac{(4\pi)^{3}}{\alpha_{G}}\simeq 5 \times 10^{4},
\ee
where we used $\alpha_{G}=1/25$. This should then be compared to the number of axions required in order to obtain a sufficient number of $e$-foldings (\ref{eq:NvsNe}). As already mentioned below Eq.~(\ref{eq:NvsNe}), the smaller the axion vevs are, the more axions are required in order to obtain sufficient amount of inflation. Therefore, if we want to obey the bound (\ref{eq:boundaxions}), the axions can not be too far down the potential when scales observed today were produced. This will have important consequences later when we calculate observables.

\item In the leading $N$-scaling, the first order string loop correction to the Planck mass is given by\footnote{In the string theory calculation there may be a correction to the standard field theory loop factor $16\pi^{2}$ \cite{Dimopoulos:2005ac}. We ignore this potential effect here but it would be worthwhile to calculate this correction term explicitly.}
\be \label{eq:gcorr}
\frac{\delta_{g_{s}}M_{Pl}^{2}}{M_{Pl}^{2}}\simeq \frac{g_{s}^{2}N}{16\pi^{2}}\frac{M_{KK}^{2}}{M_{Pl}^{2}}.
\ee
The Kaluza-Klein scale $M_{KK}$ is given by the compactification scale, which, as we have seen, is close to the string scale. Then taking $M_{KK}\lesssim l_{s}^{-1}$ allows us to write (\ref{eq:gcorr}) as
\be
\frac{\delta_{g_{s}}M_{Pl}^{2}}{M_{Pl}^{2}}\lesssim \frac{g_{s}^{2}N}{16\pi^{2}}\frac{l_{s}^{-2}}{M_{Pl}^{2}}.
\ee
We require $\frac{\delta_{g_{s}}M_{Pl}^{2}}{M_{Pl}^{2}}<1$ for perturbative control. With the string scale given by (\ref{eq:stringscale}), this gives the condition on the string coupling
\be
g_{s}^{2}<\frac{16\pi^{2}}{N}\frac{M_{Pl}^{2}}{l_{s}^{-2}}=\frac{(4\pi)^{3}}{\alpha_{G}N}.
\ee
Using the critical value for the number of axions (\ref{eq:boundaxions}), we recover the naive result
\be
g_{s}<1.
\ee
Thus, in order to maintain perturbative control, the condition on the string coupling is not sharpened by the presence of a large number of axions, as long as the bound (\ref{eq:boundaxions}) is satisfied. Taking higher loops into account the conclusion is essentially the same, assuming the $q$-loop expansion takes the general form
\be
\frac{\delta_{g_{s}}M_{Pl}^{2}}{M_{Pl}^{2}} \simeq \sum_{q}{\left(\frac{g_{s}^{2}N}{16\pi^{2}}\right)^{q}\left(\frac{M_{KK}}{M_{Pl}}\right)^{2q}}.
\ee

\item The last issue we need to consider is that of multi-instanton corrections to the potential (\ref{eq:instact2}). The general form of the decoupled potential is given by
\be \label{eq:decpot}
V^{(1)}=\sum_{n}{\Lambda_{n}^{4}(1-\cos \mbox{ }a_{n})},
\ee
where the prefactor $\Lambda_{n}$ is given by $M e^{-S_{n}/4}$. This form of the potential can be violated by multi-instanton corrections on the form \cite{Dimopoulos:2005ac,Easther:2005zr}
\be \label{eq:multipot}
V^{(2)}\sim \sum_{n,m}{\frac{\Lambda_{n}^{4}\Lambda_{m}^{4}}{M_{UV}^{4}}\cos(a_{n})\cos(a_{m})},
\ee
where $M_{UV}$ is a high energy cutoff scale. It may appear as if (\ref{eq:multipot}) is far smaller than (\ref{eq:decpot}) since the coefficients are smaller. However, since (\ref{eq:multipot}) scales like $N^{2}$, while (\ref{eq:decpot}) only scales like $N$, the situation is not entirely clear cut. In order to compare the sums we take, as before, $\Lambda_{n}=\Lambda_{m}$ for all $n$ and $m$ and assume a random distribution of the axion vevs. Then the the decoupled part of the potential is of order
\be
V^{(1)}\sim N \Lambda^{4},
\ee 
and the cross-coupling part
\be
V^{(2)}\sim \frac{N^{2}\Lambda^{8}}{M_{UV}^{4}}.
\ee
Since the decoupled part sets the inflationary scale $M_{infl}$, i.e. $V^{(1)}\sim M_{infl}^{4}$, we find when comparing these terms
\be \label{eq:fracdec}
\frac{V^{(2)}}{V^{(1)}} \sim \frac{N\Lambda^{4}}{M_{UV}^{4}} \sim \left(\frac{M_{infl}}{M_{UV}}\right)^{4} \ll 1,
\ee
for an inflationary scale ``small'' compared to the high energy cutoff scale, which must be true by consistency. In our case $M_{infl}\simeq M_{GUT}$. If we take the cutoff scale to be the string scale (\ref{eq:stringscale}), then $V^{(2)}/V^{(1)}\lesssim 10^{-3}$.  We thus conclude that cross-coupling terms can safely be ignored.

\end{itemize}

\bigskip
\section{Predictions of cosmological observables} \label{sec:cosmobs}
\bigskip

We now turn to the calculation of cosmological observables. We focus on the scalar spectral index $n_{s}$, its running with scale $dn_{s}/d\ln k$, and the tensor-to-scalar ratio $r$. These quantities are independent of the overall scale and only sensitive to the shape of the potential and the distribution of the axion vevs. As mentioned, we assume degenerate axion potentials and an initially random distribution for the axion vevs. We are, for obvious reasons, not able to exactly pinpoint the distribution of the vevs at the time when scales observed today in the CMB were produced. We deal with this by examining how the observables vary as a function of the axion vevs. Since all expressions are invariant under $a \rightarrow - a$, we take all axion vevs to be distributed in the interval $0\leq a_{i}<\pi$.

\subsection*{\it{The scalar spectral index}}

The spectral index is defined as $n_{s}-1\equiv d\ln P_{\mathcal{R}}/d \ln k$. Then $P_{\mathcal{R}}(k)\sim k^{n_{s}-1}$, so that $n_{s}=1$ implies a scale invariant power spectrum. For a single field the spectral index is given by
\be \label{eq:singspectral}
n_{s}-1=2\eta-6\epsilon,
\ee
where $\eta$ and $\epsilon$ are the standard slow-roll parameters
\be
\epsilon=\frac{M_{Pl}^{2}}{2}\left(\frac{V'}{V}\right)^{2} \qquad \eta= M_{Pl}^{2}\frac{V''}{V}.
\ee
For multiple slowly rolling and decoupled fields this generalizes to \cite{Lyth:1998xn}
\be
n_{s}-1=-M_{Pl}^{2}\left[\sum_{i}{\left(\frac{V_{,i}}{V}\right)^{2}}+\frac{2}{\sum_{i}{\left(\frac{V_{i}}{V_{,i}}\right)^{2}}}-\frac{2}{V}\frac{\sum_{i}{\frac{V_{ii}V_{i}^{2}}{V_{,i}^{2}}}}{\sum_{i}{\left(\frac{V_{i}}{V_{,i}}\right)^{2}}}\right],
\ee
which can  be seen to reduce to (\ref{eq:singspectral}) in the single-field limit. If we now plug in the potential (\ref{eq:inflpot}) and differentiate with respect to the axions, we find
\be \label{eq:explicitns}
n_{s}-1=-\frac{M_{Pl}^{2}}{F_{a}^{2}}\left[\frac{\sum_{i}{\sin^{2}a_{i}}}{\left(\sum_{i}{(1-\cos a_{i})}\right)^{2}}
+\frac{2}{\sum_{i}{\left(\frac{1-\cos a_{i}}{\sin a_{i}}\right)^{2}}}\left(1
-\frac{\sum_{i}{\frac{\cos a_{i}(1-\cos a_{i})^{2}}{\sin^{2}a_{i}}}}{\sum_{i}({1-\cos a_{i}})}\right)\right],
\ee
where the axions are now dimensionless. Note that since we have made the assumption of degenerate potentials, the prefactors of the axion potentials (\ref{eq:inflpot}) cancel out. We can write the spectral index in a more compact form by using the notation for the sums defined in Sec.~\ref{sec:axionpot}
\be \label{eq:simpns}
n_{s}-1=-\frac{M_{Pl}^{2}}{NF_{a}^{2}}\left(\frac{\bs{s_{3}}}{\bs{s_{1}}^{2}}+\frac{2}{\bs{s_{4}}}-\frac{2}{\bs{s_{1}}}\frac{\bs{s_{5}}}{\bs{s_{4}}}\right).
\ee
As a last step, using (\ref{eq:NvsNe}), we can replace the prefactor in (\ref{eq:simpns}) by the number of remaining $e$-foldings $N_{e}$
\be \label{eq:ns}
n_{s}-1=-\frac{\bs{s_{2}}}{N_{e}}\left(\frac{\bs{s_{3}}}{\bs{s_{1}}^{2}}+\frac{2}{\bs{s_{4}}}-\frac{2}{\bs{s_{1}}}\frac{\bs{s_{5}}}{\bs{s_{4}}}\right).
\ee 

Since the axion potential is quadratic close to the origin, we can, as a consistency check, verify that (\ref{eq:ns}) reduces to the standard result for $m^2\phi^2$-inflation in the limit of small and equal fields
\be
n_{s}-1\big{|}_{a_{i}=a_{j}\rightarrow 0} \rightarrow -\frac{2}{N_{e}}.
\ee

As is clear from (\ref{eq:ns}), the spectral index only depends on how the axions are distributed at the time when $N_{e}$ $e$-foldings remains of inflation. Initially we assume a completely random distribution, which is quite reasonable given that we have of order $10^{4}$ axions with independently broken shift-symmetries. The question then is how far down the potentials the axions have reached when $N_{e}$ $e$-foldings remains of inflation. Without knowing this exactly, the best thing we can do is to express the spectral index as a function of a parameter measuring the average vevs.

\begin{figure}[t]
\centering
\includegraphics[height=8cm, width=10cm]{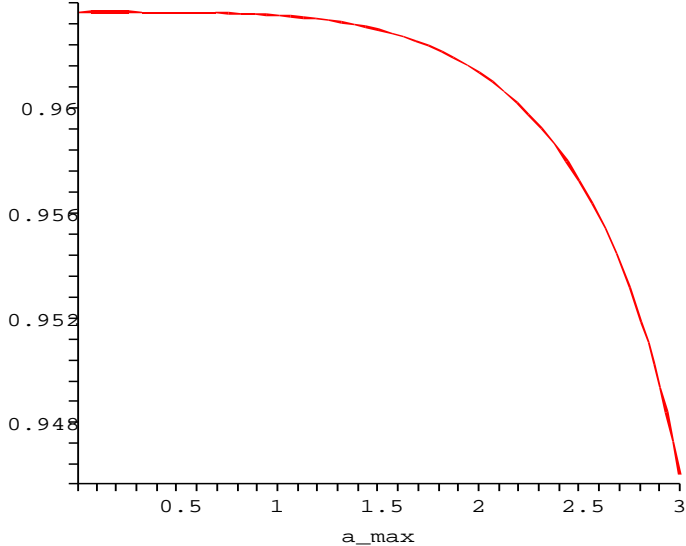}
\caption{The spectral index $55$ $e$-foldings from the end of inflation for degenerate axion potentials, as a function of the maximal axion vev $a_{max}$. It is assumed that the other axion vevs are equidistantly distributed in the interval $0\leq a_{i}<a_{max}$.}
\label{fig:nsfunc}
\end{figure}

We take this parameter to be the axion with greatest vev $a_{max}$. For simplicity, we assume the other axions to be equidistantly distributed in the interval $0\leq a_{i}<a_{max}$.\footnote{This is a quite reasonable assumption at early stages of inflation, but less so at later stages when the axions start to crowd at the bottom of the potential.} In this way the sums in (\ref{eq:ns}) become functions of the parameter $a_{max}$. We can then plot the spectral index (\ref{eq:ns}) as a function of this parameter. The result is shown in Fig.~\ref{fig:nsfunc}.

The spectral index decreases monotonically with $a_{max}$. Note that for small axion vevs, the spectral index approaches the result for $m^2\phi^2$-inflation, i.e. $n_{s}=1-2/N_{e}=0.963$. However, since we expect that initially the axion vevs are evenly distributed in the interval $0\leq a<\pi$, inflation will not be maintained for another $55$ $e$-foldings when the axions are this far down the potential.

At the other end of the spectrum, where the axions are placed equidistantly in the interval $0\leq a<3$, we find $n_{s}=0.946$, a result which substantially differs from $m^{2}\phi^{2}$-inflation. This would be the result if inflation did not last much longer than $55$ $e$-foldings.\footnote{The reason we do not include axions at the top of the potential, $a_{max}=\pi$, is that these axions will stay there forever as long as they are not perturbed. Quantum fluctuations will, however, always perturb axions from this point. Therefore, we do not need to worry about axions in this regime soon after inflation begins. For definiteness, we assume that $a_{max}\leq3$, when scales observed today were produced.} The number of axions required to get at least $55$ $e$-foldings of inflation with this distribution is given by (\ref{eq:NvsNe}), which gives
\be \label{eq:Nbound}
N\gtrsim\frac{2N_{e}S^{2}}{\bs{s_{2}}}\simeq 1.3 \times 10^{4},
\ee
where $S=18.3 g_{s}^{2/3}$ and $g_{s}\simeq 0.5$. This number is safely below the upper bound for the number of axions (\ref{eq:boundaxions}).

The further down the potential the axions are, the more axions are needed to get the required number of $e$-foldings. We can use the axion bound (\ref{eq:boundaxions}) to find the lowest acceptable $a_{max}$ in order to get $55$ $e$-foldings without having to worry about higher order corrections. Then from (\ref{eq:NvsNe}) it is found that $a_{max} \simeq 1.8$ is the critical lower value for the axion with greatest vev, for the bound (\ref{eq:boundaxions}) to be satisfied. For this critical case $n_{s}\simeq 0.962$. We thus conclude that within the regime where the leading order theory is valid, the spectral index lies in the interval $0.946\lesssim n_{s}\lesssim 0.962$. This should be compared to the $3$rd year WMAP result \cite{Spergel:2006hy} $0.932<n_{s}^{obs}<0.966$.

The upper bound for the spectral index can be further sharpened if it is found that the maximal number of axions attainable in compactifications of heterotic string theory is less than the upper bound (\ref{eq:boundaxions}). If, say, there are at most $2\times 10^{4}$ axions, then, in order to get at least $55$ $e$-foldings of inflation, (\ref{eq:NvsNe}) requires that $a_{max}>2.6$. This would result in $n_{s}<0.956$. The theoretic gap would then only be $\delta{n_{s}}\lesssim 0.010$.

\subsection*{\it{Running of the spectral index}}

Let us now investigate how the spectral index runs with scale, i.e. $dn_{s}/d \ln{k}$. For decoupled fields we can use the general formula given in \cite{Lyth:1998xn}. The derivative can be rewritten as
\be
\frac{d}{d\ln{k}}=-\frac{M_{Pl}^{2}}{V}\frac{\partial V}{\partial a^{i}}\frac{\partial}{\partial a_{i}}=
-\frac{M_{Pl}^{2}}{F_{a}^{2}N\bs{s_{2}}}\sum_{i}{\sin{a_{i}}\frac{\partial}{\partial a_{i}}}.
\ee 
Applying this to the spectral index (\ref{eq:ns}) gives
\be \label{eq:runningns}
\frac{dn_{s}}{d \ln{k}}=
-\frac{2\bs{s_{2}}^{2}}{N_{e}^{2}\bs{s_{1}}}\bs{S},
\ee
where
\be
\bs{S}=\frac{1}{N} \sum_{i}{\sin^{2} a_{i}\left[\frac{\cos a_{i}}{\bs{s_{2}}^{2}}+\frac{\bs{s_{3}}}{\bs{s_{1}}^{3}}+\frac{\bs{s_{5}}}{\bs{s_{1}}^{2}\bs{s_{4}}}+\frac{2\bs{s_{5}}-2\bs{s_{1}}+\bs{s_{4}}(1-\cos^{2}a_{i}-2\cos{a_{i}})}{\bs{s_{4}}^{2}\bs{s_{1}}(1+\cos a_{i})^{2}}\right]}.
\ee
As a consistency check, it is straightforward to verify that (\ref{eq:runningns}) reduce to the standard $m^{2}\phi^{2}$-result
\be
\frac{dn_{s}}{d \ln{k}}\rightarrow -\frac{2}{N_{e}^{2}},
\ee
in the limit of small and equal fields. From (\ref{eq:runningns}) one can also verify the inequality
\be
\left|\frac{dn_{s}}{d \ln{k}}\right| \leq \frac{2}{N_{e}^{2}}=6.6\times 10^{-4}.
\ee 
Since the running of the spectral index in $m^2\phi^2$-inflation is already too small to be detected, we conclude that it is completely negligible in this model.

\subsection*{\it{Tensor perturbations}}

We next turn to tensor perturbations. These are fluctuations of the background itself. The power spectrum of these fluctuations is directly given by the inflationary potential $V$ according to
\be \label{eq:deftensorp}
P_{g}=\frac{2}{3\pi^{2}}\frac{V}{M_{Pl}^{4}},
\ee
where we used the convention of \cite{Peiris:2003ff}.

A convenient measure of the tensor contribution $P_{g}$ to the CMB power spectrum is its relative size to the scalar part $P_{\mathcal{R}}$. This is the the so called tensor-to-scalar ratio $r \equiv P_{g}/P_{\mathcal{R}}$. From (\ref{eq:defscalarpower}) and (\ref{eq:deftensorp}), we find
\be \label{eq:r}
r\equiv \frac{P_{g}}{P_{\mathcal{R}}}=\frac{8M_{Pl}^{2}}{\sum_{i}{\left(\frac{V_{i}}{V_{,i}}\right)^{2}}}=\frac{8M_{Pl}^{2}}{F_{a}^{2}N\bs{s_{4}}}=\frac{8}{N_{e}}\frac{\bs{s_{2}}}{\bs{s_{4}}},
\ee
where in the last step we used Eq.~(\ref{eq:NvsNe}), and the sums $\bs{s_{2}}$ and $\bs{s_{4}}$ were defined in (\ref{eq:sum2}) and (\ref{eq:sum4}).
Again, as a consistency check, we verify that (\ref{eq:r}) reduce to the standard $m^{2}\phi^{2}$-result $r\rightarrow 8/N_{e}$, in the limit of small and equal fields. This is also the greatest possible value of $r$.

\begin{figure}[t]
\centering
\includegraphics[height=8cm, width=10cm]{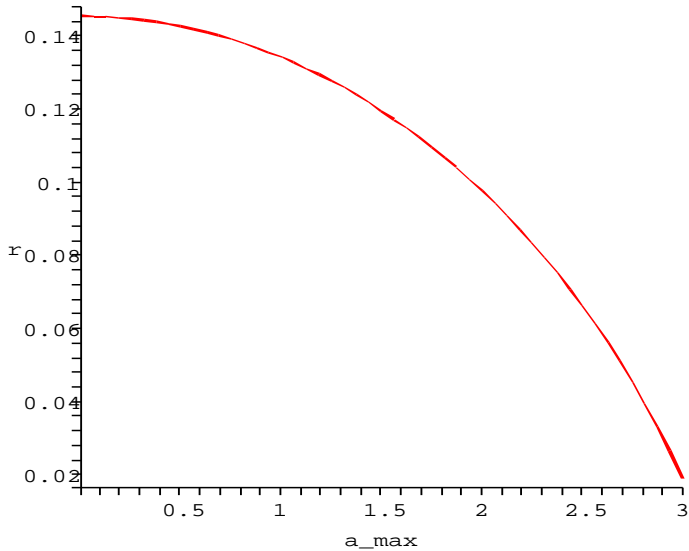}
\caption{The tensor-to-scalar ratio $55$ $e$-foldings from the end of inflation for degenerate axion potentials, as a function of the maximal axion vev $a_{max}$. It is assumed that the other axion vevs are equidistantly distributed in the interval $0\leq a_{i}<a_{max}$.}
\label{fig:rfunc}
\end{figure}
As for the scalar spectral index, we display the general result for the tensor-to-scalar ratio as a function of the greatest axion vev $a_{max}$, and take the other axions to be equidistantly distributed in the interval $0 \leq a_{i} <a_{max}$. The result is shown in Fig. \ref{fig:rfunc}.

Clearly, the tensor-to-scalar ratio is much more sensitive to axion vevs than the spectral index. It is generically smaller than for $m^{2}\phi^{2}$-inflation, and substantially smaller already for moderate axion vevs. The underlying reason is that the scalar power spectrum has a dependence $P_{\mathcal{R}} \sim V\sum{(V_{i}/V_{,i})^{2}}$ (\ref{eq:scalarpower}), where $\sum{(V_{i}/V_{,i})^{2}}$ grows with the axion vevs. Then for larger vevs, in order to keep $P_{\mathcal{R}}$ fixed, the potential $V$ must be decreased. Hence, since the tensor power spectrum is given by the potential alone, the tensor-to-scalar ratio becomes smaller for larger axion vevs.

In the discussion of the spectral index we found that in order to trust the leading order theory (i.e.~obey the bound (\ref{eq:boundaxions})), and obtain at least $55$ $e$-foldings of inflation, we must require $a_{max}>1.8$. For the tensor-to-scalar ratio this implies (\ref{eq:r})
\be
r<0.10,
\ee
for the regime where the leading order theory is valid. As for the spectral index, this bound can be further sharpened for compactifications with less axions than the upper bound (\ref{eq:boundaxions}). If we take again $N=2 \times 10^{4}$, then we must require $a_{max}>2.6$ in order to get at least $55$ $e$-foldings of inflation (\ref{eq:NvsNe}). It then follows from (\ref{eq:r}) that the tensor-to-scalar ratio is bounded by $r<0.058$.

Note, however, that $r$ is large enough to be detected by future observations -- see for example \cite{Ungarelli:2005qb} for estimated sensitivity of upcoming observations. If we assume the greatest vev was no larger than $a_{max}\simeq 3$ when scales observed today were produced, we get $r\gtrsim 0.02$, which should be large enough to be detected by future observations.

It is also straightforward to cross-correlate the scalar spectral index (\ref{eq:ns}) with the tensor-to-scalar ratio (\ref{eq:r}). For example, an observation of $n_{s} \simeq 0.951$ corresponds, in this model of inflation, to $r \simeq 0.036$.

\bigskip
\section{Concluding remarks} \label{sec:conclusions}

We have presented an attempt to explicitly realize N-flation in heterotic string theory, where higher order corrections can be kept under control and no fine-tuning is required. The observables in this model come out within present observational bounds. Interestingly, requiring that higher order effects are small compared to leading order results imposes rather strong constraints on the observables in question.

The results of the leading order theory may nevertheless be modified in a more careful treatment than the one presented here. In particular, deviations from degenerate potentials (if there are any) will make the spectral index more red \cite{Easther:2005zr}. Furthermore, since we are dealing with many inflaton fields with independent and random initial vevs, later stages of inflation should be treated beyond the first order slow-roll approximation. Also, the actual axion distribution, and its subsequent evolution will be more complicated than treated here. For nondegenerate potentials, it will be substantially more complicated.

Let us make a few more comments on the assumption we made for the axion potentials. The sizes of the potentials are given by the instanton actions. These instanton effects are due to fundamental strings wrapping two-cycles in the compactification manifold. The actions are then proportional to the size of these two-cycles. As mentioned in the main text, it is encouraging that the compactification volume is so close to the string scale (\ref{eq:hetvol}), $V_{z}\simeq 5 l_{s}^{6}$ with $g_{s}\simeq 0.5$, which implies that the internal two-cycles cannot vary much in size. However, the potentials have an exponential dependence on the instanton actions, which means that modest variations of the sizes of the two-cycles may lead to rather large variations of the potentials.

As has been mentioned, the general case for small initial vevs and nondegenerate axion potentials has been analyzed in \cite{Easther:2005zr}. We have treated random initial vevs and degenerate potentials in the specific case of the heterotic string. It would be interesting to combine these two results in heterotic string theory.

We focused on heterotic string theory since in this case we found that it was possible to get a potential large enough to obtain inflation which generates the correct order of magnitude for the CMB fluctuations. This was directly related to the fact that the instantons generating the potential are world-sheet instantons. The heterotic string may not be the only string theory which satisfies the required constraints, and it would be useful to explore other string models. However, due to the way the strength of the instanton effects depends on the string coupling, at appears that at weak coupling, D-brane instantons generates potentials too small for this purpose. If this is true in general, many string theory models are ruled out. 

One important issue in inflationary cosmology is the process of reheating, where the inflaton transfers its energy to radiation and matter fields. Since we are here dealing with inflatons protected by a (discrete) shift-symmetry, the coupling to other fields is quite weak. Thus, it would be interesting to investigate what the implications are regarding the reheat process, when using axions as inflaton fields.

In heterotic string theory there is also the model independent axion. Its influence during inflation is completely negligible since its potential is very small compared to the model dependent ones, roughly one order of magnitude smaller in the exponent \cite{Svrcek:2006yi,Svrcek:2006hf}. However, because of its small mass and weak coupling to ordinary matter, it may be important in late time cosmology as it contributes to dark matter. It would be interesting to explore its evolution and resulting contribution to the matter density at later stages of the universe.

We conclude this paper by a speculative remark. We have investigated the \emph{weakly} coupled heterotic string in order to explain the cosmic acceleration in the \emph{early} universe. On the other hand, it has been observed that the model dependent axions in \emph{strongly} coupled heterotic string theory, i.e. heterotic M-theory, have potentials with the correct order of magnitude in order to account for the \emph{currently} observed vacuum energy \cite{Svrcek:2006hf}. In this regard, we find it intriguing that -- as was realized very recently \cite{Hellerman:2006nx,Hellerman:2006ff,Hellerman:2006hf} -- explicit examples of cosmological-like transitions between different string theories, do exist.

\bigskip
\section*{Acknowledgments}

I would like to thank Anupam Mazumdar for discussions and Kristjan R. Kristjansson for helpful comments on the manuscript.

\bibliographystyle{utcaps}
\bibliography{mybibfile}

\end{document}